\documentclass[aps,preprint,aps,12pt]{revtex4}%
\usepackage{times}
\usepackage{amsmath}
\usepackage{epsfig}
\renewcommand{\baselinestretch}{2}

\usepackage{graphicx}

\begin{document}

\title{\bf
Structural and magnetic properties of Mn-doped GaAs(110) surface}
\newpage
\begin{abstract}
We have investigated STM images of the (110) cross-sectional
surface of Mn-doped GaAs  using first principles total-energy
pseudopotential calculations. We focus on configurations with Mn
interstitial in the uppermost surface layers. In particular, we
have found that Mn impurities, surrounded by Ga or As atoms,
introduce in both cases strong local distortions in the GaAs(110)
surface, with bond length variations up to 8\% on surface and
non-negligible relaxations effects propagating up to the third
sub-surface layer. In both cases interstitial Mn induces a
spin-polarization on its nearest neighbors, giving rise to a
ferromagnetic Mn--As and to antiferromagnetic Mn--Ga
configuration.

\end{abstract}

\author{A. Stroppa}
\author{X. Duan}

\affiliation{Dipartimento di Fisica Teorica, Universit\`a di
Trieste,\\ Strada Costiera 11, I-34014 Trieste, Italy}
\affiliation{INFM DEMOCRITOS National Simulation Center, Trieste,
Italy}

\author{M. Peressi}
\affiliation{Dipartimento di Fisica Teorica, Universit\`a di
Trieste,\\ Strada Costiera 11, I-34014 Trieste, Italy}
\affiliation{INFM DEMOCRITOS National Simulation Center, Trieste,
Italy}

 \maketitle

\newpage
\renewcommand{\baselinestretch}{2}
\renewcommand{\thesection}{\arabic{section}}

 \section{Introduction}

 Diluted magnetic semiconductors (DMS's) have
been considered of tremendous scientific and technological
importance.\cite{spintronic1,spintronic2,spintronic3} This is
essentially due to the combination of ferromagnetism with
semiconducting properties in the same host material which enable
the use of the spin degree of freedom to process, to transfer as
well as to store information,
giving rise to the emerging field of \emph{Spintronic}.\cite{Spinbis}\\
Among DMS's materials, ferromagnetic
 (Ga,Mn)As has attracted considerable attention. Substitution of Mn for Ga
in GaAs introduces a local spin $\frac{5}{2}$ magnetic moment, and
acts as an acceptor,
 providing itinerant holes which mediate the ferromagnetic order.\cite{gamnas1}
An important step toward near future device applications  was
achieved some years ago, when it was recognized that annealing at
temperatures close to the growth temperature can result in an
important improvement of the Curie temperature (the highest
T$_{C}$ for the past few years was 110 K). The observed changes
have been attributed to out diffusion of Mn \emph{interstitials}
towards the surface.\cite{PRLMN} Therefore, it is of great
importance for practical applications to clearly understand the
role of  Mn-dopant in determining the
magnetic and electronic properties.\\
It is clear that the properties of such systems strongly depend on
the type and concentration of defects.\cite{PRLMN} In this
perspective, experimental and theoretical studies of the
atomic-scale structure of (Ga,Mn)As are highly motivated.\\
From the experimental point of view, Mn $\delta$-doped GaAs
samples have been recently grown in (001) direction at TASC
Laboratory in Trieste.\cite{Modesti} The cleavage of these samples
along the natural (110) cleavage plane yields large automatically
flat surfaces with Mn dopants on or close to the exposed surface,
thus allowing to study Mn defects environment with
surface sensitive techniques.\\
In this context, {\it Cross-sectional Scanning Tunnelling
Microscopy} (XSTM) is  a powerful tool. With the purpose of
characterizing the local environments of defects, we have
simulated XSTM images for different Mn configurations and compared
with available experimental images. We focus our attention here on
the impurity
interstitial surface configurations.\\
This paper is organized as follows: in the next section we describe
the computational method; in Sect. 3 we present our results for the
structural and magnetic properties; in Sect. 4 we discuss the XSTM
images; finally, in Sect. 5 we draw our conclusions.

\section{Computational details}
Our calculations have been performed within Density Functional
Theory (DFT) framework in the Local Density Approximation for the
exchange-correlation functional\cite{PZ,PZ1}, using
state-of-the-art first-principles pseudopotential self-consistent
calculations, as implemented in the ESPRESSO/PWscf
code\cite{pwscf}. Ultrasoft (US) pseudopotential (PP)\cite{Vander}
has been used for Mn atom, while norm-conserving PPs have been
used for Ga, As and H atoms. Test calculations have shown that a
kinetic energy cutoff for the wave functions equal to 22 Ry and a
200 Ry cutoff for the charge density are sufficient to get well
converged results. We
  estimate the numerical uncertainty  to be
  $\sim$ 0.001 nm for relative atomic displacements and
 $\sim$  0.01 $\mu_{B}$ for the magnetic moments.
The relaxed internal atomic positions
 have been obtained by total-energy and atomic-force minimization using the
 Hellmann-Feynman theorem.\cite{forces}\\
 We model the
  surface using the supercell approach, with
periodically repeated cell containing one Mn atom; a (110) slab
geometry with
 a 4$\times4$ in-plane periodicity has been used. The simulation cells are
made up of 5 atomic layers and a vacuum region equivalent to 8
atomic layers. The bottom layer has been passivated with Hydrogen
atoms. Only the three uppermost layer are allowed to relax, while
the others are kept fixed. Two different configurations have been
considered for Mn on the surface, namely Int$_{Ga(As)}$ (see next
Section). In each case, the distances between the Mn atom and its
periodic image on the (110) plane are 1.57 nm along the
[1$\bar{1}$0] and 2.22 nm along [001]: test calculations demonstrate
that the supercell is large enough to
neglect the Mn-Mn interactions.\\
XSTM images are simulated using the model of
 Tersoff-Hamann\cite{Ters1,Ters2}, where a point-like tip is assumed
 and the tunneling current is derived from the local density of states at the Fermi energy, E$_{f}$.
 Within this approximated model, the constant current STM images are
 simulated from electronic structure calculations by considering
 surfaces of constant local density of states integrated over an
 energy window from E$_{f}$ to E$_{f}$+V, where V is the voltage applied
 between the sample and the tip. In
 this model the tip near the surface does not influence the electronic states.\\

\section{Structural and magnetic properties}
\renewcommand{\thesubsection}{\arabic{subsection}}
\subsection{Structural properties}
The GaAs(110) relaxed structure is well known from experimental as
well as theoretical point of view. In the relaxed surface, the
electronic charge is transferred from Ga to As atoms with the
occupied state density being  localized around surface As atoms
and the unoccupied density around the Ga atoms.\cite{STMGAAS} This
charge transfer is accompanied by an approximately
bond-length-conserving rotation with As atoms moving upward and Ga
atoms moving downward, still preserving the 1$\times$1 bulk
periodicity. Due to overbinding in the  LDA approximation, our
theoretical GaAs lattice constant (0.555 nm) is smaller than the
experimental one (0.565 nm) but the relevant calculated structural
parameters for the clean surface
 such as $\Delta_{1,\bot}$ (relative displacement of the anion and
 cation positions in the uppermost layer, normal to the surface) and $\alpha$ (the
buckling angle)
  are 0.068 nm and 30.36$^{\circ}$ respectively, which well compare with the
experimental values 0.065 nm and 27.4$^{\circ}$.\cite{Duke} The
clean surface remains
 semiconducting with a calculated energy gap $\sim$ 0.72 eV.\\
Throughout this work, we have considered only \emph{tetrahedral}
interstitial position, as the total energy corresponding to the
\emph{hexagonal} interstitial one is higher by more than 0.5
eV.\cite{bulkint,PRLMN,condmat} In the bulk zinc-blende crystal
structure, there are two inequivalent tetrahedral interstitial
position  which differ in the local environment. We call them
Int$_{Ga(As)}$, to denote that  Mn  is surrounded by four Ga(As)
atoms. The tetrahedral interstitial positions in the ideal
geometry is equidistant from its four nearest-neighbor (NN) atoms
with a distance equal to the ideal host bond length $d_{1}$. There
are six next-nearest-neighbor (NNN) atoms at the distance
$d_{2}=\frac{2}{\sqrt{3}}d_{1}$, which are
 As(Ga) atoms for Int$_{Ga(As)}$, respectively.\\
At surface, the tetrahedral interstitial position has three NNs
and four NNNs instead of four and six respectively as in the bulk
case. To start with, we consider the \emph{clean} and
\emph{relaxed} GaAs(110) surface with the Mn position such that
the NN bond lengths are all equal. This configuration  will be
referred to as \emph{initial} in the following. Due to symmetry
breaking because of the surface and the consequent buckling of the
outermost surface layers, the NNN bond lengths are
no longer equal.\\
After relaxations, the two configurations, Int$_{Ga}$ and
Int$_{As}$, are almost degenerate, differing by $\sim$ 130 meV/Mn
atom
(Int$_{Ga}$ is favoured).\\
 In Fig.\ref{Fig1} we show a ball and stick side (a) and
top (b) view of the relaxed Int$_{Ga}$ and Int$_{As}$
configurations. Only the three topmost layers and the atoms
closest to Mn are shown. Grey spheres are cations (Ga atoms),
white spheres are anions
 (As atoms); Mn is explicitly indicated. It is easy to see that
 the presence of Mn strongly reduces the surface buckling.
 In Fig.\ref{Fig1}b,
 atomic moments are also indicated for atoms close to Mn and
  the numbers in parenthesis specify the atomic
 layer from the surface. To characterize the relaxed configurations,
in Table \ref{tab1} we report the NN and NNN bond lengths in the
\emph{relaxed} and \emph{initial} (in square brackets)
configurations. The atomic types are in round brackets and 1$^{st}$  and 2$^{nd}$ denote
the two uppermost layers.\\
For Int$_{Ga}$, the two surface Mn-Ga bonds increase by $\sim$ 4.6
\%, from 0.237 to 0.248 nm, whereas the backbond to the Ga atom in
the layer beneath increase by $\sim$ 8.0 \% (from 0.237 to 0.256
nm). The NNN bond-lengths relaxations are less pronounced with
elongations of about $\sim$ 2-5 \%. In the other configuration, the
two  surface bonds between Mn and As elongate by 2.2 \% from 0.247
nm to 0.252 nm whereas the bond with subsurface As
shrinks by 1.2 \% (from 0.247 nm to 0.244 nm). The relaxations leave almost unchanged
the NNN bond lengths when Ga atom belongs to 2$^{nd}$ layer whereas the surface interatomic
Mn--Ga distance is strongly reduced with respect to the initial one.
 From Fig. \ref{Fig1}a, we see that small relaxations effects are
still present in the third layer, in both configurations.\\
In conclusion, the largest local distortions with respect to the
\emph{clean surface} occur in the Int$_{Ga}$ configuration
resulting in a remarkable repulsion of the NNs and NNNs whereas,
in Int$_{As}$, the  lattice relaxations around the Mn
impurity involve mainly the NNN Ga atom on surface.

\subsection{Magnetic properties}

In the following, we analyze the magnetic properties for the two
configurations. In Fig.\ref{Fig1}b (top views), we report the
spin-polarizations for Mn and for the NN and NNN atoms. The
highest value of Mn spin-polarization is found in Int$_{As}$, with
$\mu_{Mn}$=3.96
 $\mu_{B}$. In the other configuration, the
 Mn magnetic moment is 3.67 $\mu_{B}$.
 From the angular-momentum, spin- decomposed charge,
one recognizes that the Mn magnetic moment mostly derives  from
$d$ polarization while the $s$-Mn states are only slightly
  polarized with 0.07(0.05)  $\mu_{B}$ for Int$_{Ga(As)}$.
  The total and absolute magnetization in the supercell are
  different in both cases. This corresponds to the presence of
  antiferromagnetic regions coupled to Mn. The total and absolute magnetization
  are 3.41(4.23) and
  4.71(4.84) $\mu_{B}$  for Int$_{Ga(As)}$
  respectively thus suggesting  that the region of negative
  magnetization should be larger in Int$_{Ga}$ respect to
  Int$_{As}$.\\
Let us focus on Int$_{Ga}$ configuration. The two surface Ga NN of
Mn have an induced polarization opposite to Mn magnetic moment,
equal to -0.17 $\mu_{B}$, mostly due $p$ polarization (induced
through hybridization with $d$ states); the other Ga atoms have a
negligible polarization. The induced polarization on surface As
atoms are negligible, while it is equal to 0.05
 $\mu_{B}$ for the atom on the 2$^{nd}$ layer.\\
For Int$_{As}$, the NN As atoms show a ferromagnetic coupling with
Mn (see Fig.\ref{Fig1}b), with a magnetic moment equal to 0.05
$\mu_{B}$. The induced polarization in more distant As atoms is
strongly reduced although non negligible up to the  fourth-layer
As atom. When considering the Ga atoms around the Mn, we see an
antiferromagnetic coupling between Mn and  surface Ga atom,
  with an enhanced polarization compared to As (the Ga moment is
0.14 $\mu_{B}$). The
 polarization on the
other closest Ga atoms is negligible.\\
Our results for the magnetic properties can be summarized as
follows: in both cases, the surface Ga atom(s) close to Mn are
coupled antiferromagnetically, whereas those subsurface have
negligible spin-polarization; the As atoms are in all cases
coupled ferromagnetically to Mn, with spin-polarization on the
surface as well as on subsurface atoms.

\section{STM Images}
We show the schematic front and side views  of the relaxed
underlying structure lattice and the XSTM images, with the actual
size ($\approx$ 2.2 nm $\times$ 1.6 nm) of the supercell used in
the simulations, at negative and positive bias voltages (from V =
$-$ 2.0 V to $+$2.0 V). In the simulated images, the $E_{f}$ is
near the Valence Band Maximum (VBM), in order to simulate the
experimental conditions of $p$-doped samples.\cite{Modesti}

\subsection{Isolated Mn Interstitial (Int$_{Ga}$)}

In Fig. \ref{intga}, we show the simulated STM images for the
isolated Mn in the Int$_{Ga}$ relaxed configuration. A dark region
appears around Mn atom at filled states. At positive bias
voltages, the two NN surface Ga atoms of Mn appear very bright
with features extending towards the Mn and the atoms in the
neighbourhood also looking brighter than normal. These features
change a little bit according to the specific positive bias
applied, but do not disappear.
\subsection{Isolated Mn Interstitial (Int$_{As}$)}

In Fig.\ref{intas}, we show the simulated XSTM images for Mn in
Int$_{As}$ configuration.
 At
 negative bias Mn  appears as an additional bright spot close to its
 neighbouring surface As
 atoms. If we change V from -1 to -2
V this feature remains but it is attenuated. A very bright
elongated spot in the center of the surface unit cell delimited by
As is visible at positive bias voltage which is contributed mainly
by Mn atoms, specifically by Mn 3d spin-up electron and surface Ga
empty states. When we increase the bias voltage to 2 V, the
interstitial Mn atom still appears brighter. For this case, the
simulated XSTM images show common features with the experimental
images.\cite{Modesti}

\section{Conclusion}
In summary, we have used first-principles simulations to
characterize Mn interstitial impurity on the GaAs(110) surface.
From total energy calculation, Int$_{Ga}$ and Int$_{As}$ are
almost degenerate in energy. Strong local distortions on the (110)
GaAs surface are introduced by Mn, especially when it is
surrounded by Ga atoms. Small relaxations effects are found up to
the third sub-suface layer. In both case, Mn polarizes the NN and
NNN atoms, giving rise to a ferromagnetic Mn--As and to an
antiferromagnetic Mn--Ga configuration. Comparison of simulated
XSTM images with experimental ones preliminary available seem to
indicate an Int$_{As}$ configuration in the experimental samples.

\section{Acknowledgments}
The authors would like to thank S. Modesti and D. Furlanetto for
fruitful discussions. Computational resources have been obtained
partly within the ``Iniziativa Trasversale di Calcolo Parallelo''
of the Italian {\em Istituto Nazionale per la Fisica della
Materia} (INFM) and partly within the agreement between the
University of Trieste and the Consorzio Interuniversitario CINECA
(Italy).

\renewcommand{\baselinestretch}{1}

{}

\newpage

\renewcommand{\baselinestretch}{2}

\clearpage

\begin{figure}[!hbp]

\caption{Schematic side (a) and top
 (b) views of the relaxed Int$_{Ga}$ (left) and Int$_{As}$ (right)
 configurations.
 Only the three topmost layers are shown.
 Grey spheres are cations (Ga atoms), white spheres are anion
 (As atoms),  Mn is explicitly shown. Atomic moments for atoms close
 to Mn are shown in
 the top views (b) and the numbers in brackets specify also the atomic
 layer, when not evident. Units are in $\mu_{B}$.}\label{Fig1}
\end{figure}

\vspace{1cm}

\newpage

\begin{figure}[!hbp]

\caption{Simulated STM images of isolated Mn interstitial in
GaAs(110) surface, with Ga NNs (Int$_{Ga}$). Top panel: ball and
stick model of the relaxed surface, top and side view (Ga: empty
circle, As: filled circle, Mn: square). Bottom panels: simulated
STM images at occupied states and empty states respectively, for
different bias voltages.}\label{intga}
\end{figure}

\vspace{1cm}
\newpage
\begin{figure}[!hbp]

\caption{Simulated STM images of isolated Mn interstitial in
GaAs(110) surface, with As NNs: (Int$_{As}$ in the text). See
caption of Fig. 2 other details.}\label{intas}
\end{figure}

\clearpage
\begin{table}[!hbp]
\caption{Nearest-neighbor (NN) and next-nearest neighbor (NNN)
bond-lengths for relaxed Int$_{Ga}$ (upper part) and Int$_{As}$
(lower part); 1$^{st}$ and 2$^{nd}$  refer to the atomic layer from the
surface and the kind of atoms bonded to Mn (See Fig. 1) are in round
brackets; the numbers in square brackets refer to \emph{initial}
bond lengths (see text). Units are in nm.}\label{tab1}

\vspace{1.cm}
\newpage

\begin{tabular}{||c|c|c|c||}

\hline\hline

\hline

\multicolumn{2}{|c|}{NN(nm)}&
\multicolumn{2}{|c||}{NNN(nm)}\\
\hline
\multicolumn{4}{||c||}{Int$_{Ga}$}\\
\hline 1$^{st}$(Ga) &2$^{nd}$(Ga) &1$^{st}$(As)
&2$^{nd}$(As) \\
\hline

\hline

0.248[0.237]&0.255[0.237]&0.263[0.254]&0.268[0.257]\\

\hline\hline
\multicolumn{4}{||c||}{Int$_{As}$}\\
\hline
1$^{st}$(As)&2$^{nd}$(As)&1$^{st}$(Ga)&2$^{nd}$(Ga)\\

\hline
0.252[0.247]&0.244[0.247]&0.249[0.298]&0.290[0.291]\\

\hline\hline

\end{tabular}
\end{table}

\clearpage

\clearpage

\begin{figure}[!hbp]
\includegraphics[scale=.8,angle=0.]{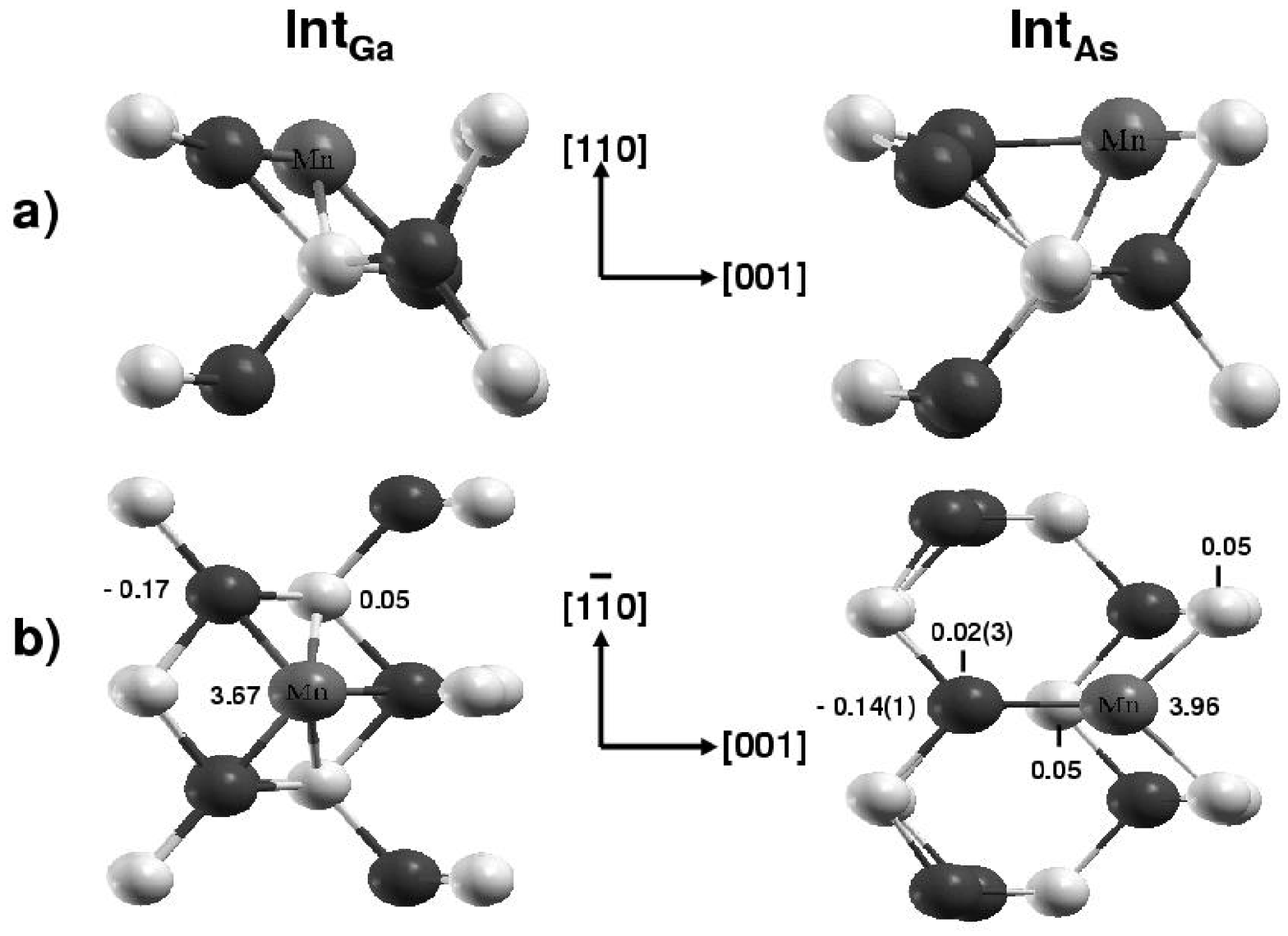}
\\Fig. 1
\end{figure}

\clearpage

\begin{center}
\begin{figure}
~~~~~~~~\includegraphics[angle=0,width=.49\textwidth]{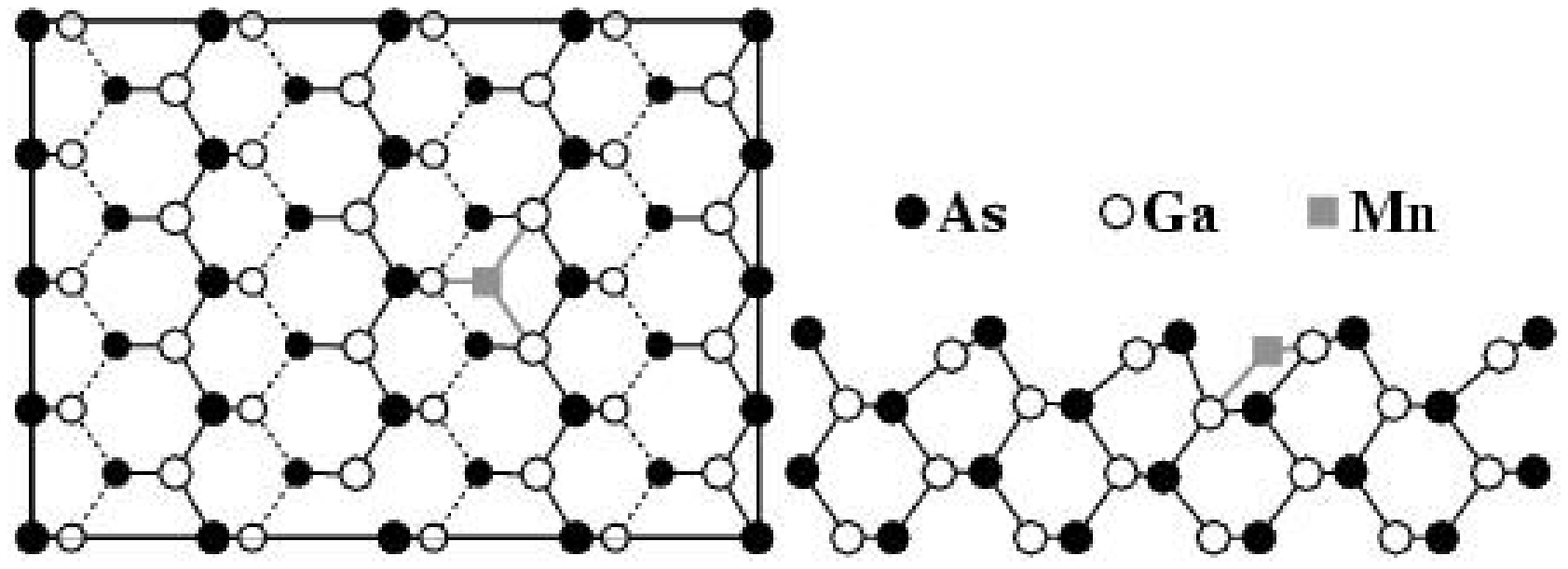}\\
\vspace{0.5cm}
\includegraphics[angle=0,width=.42\textwidth]{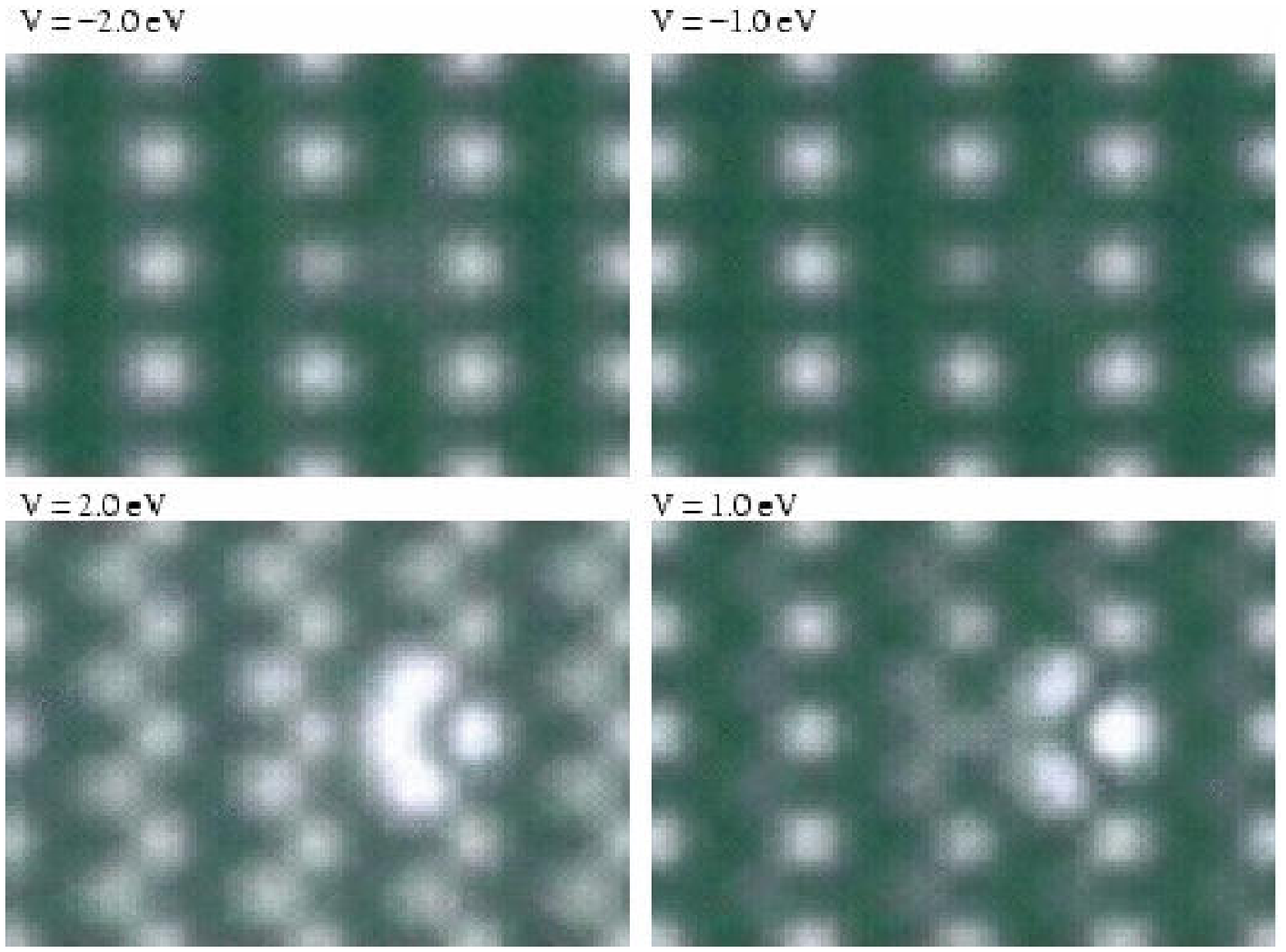}
\\Fig. 2
\end{figure}
\end{center}

\newpage
\clearpage

\begin{figure}
\vspace{2.5cm}
~~~~~~~\includegraphics[angle=0,width=.5\textwidth]{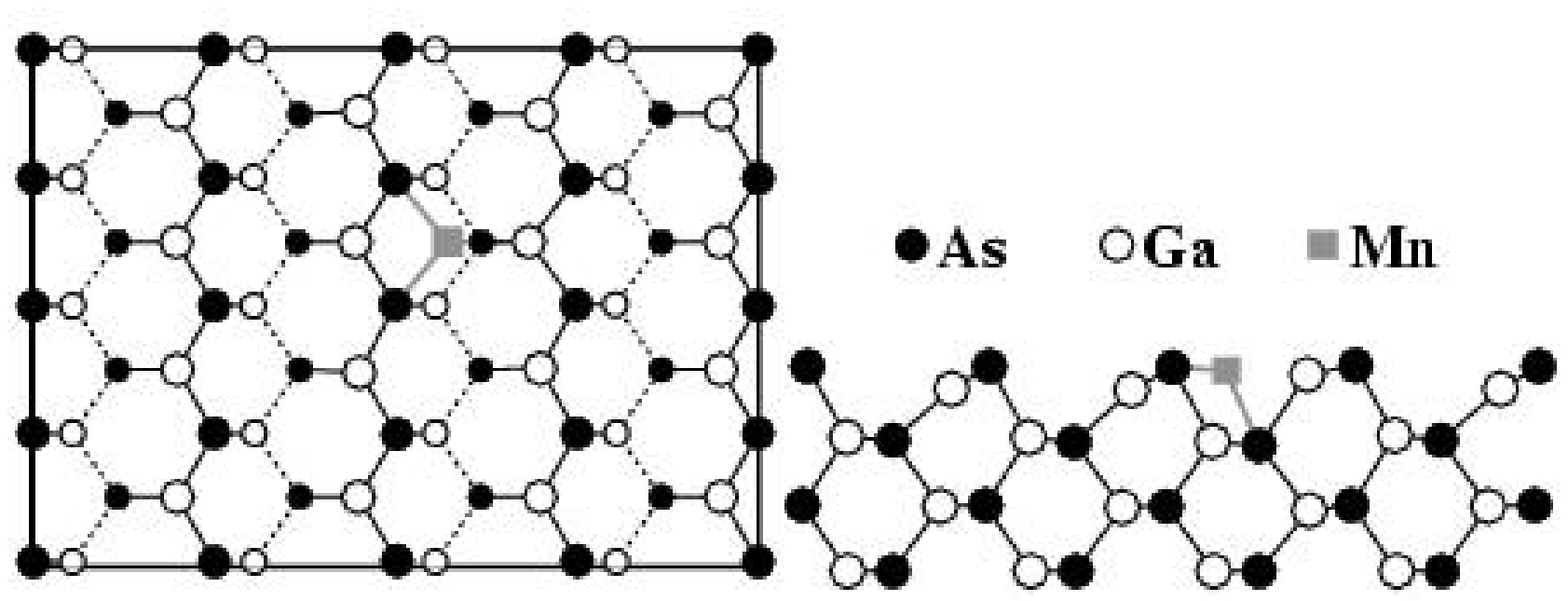}\\
\vspace{0.5cm}
\includegraphics[angle=0,width=.42\textwidth]{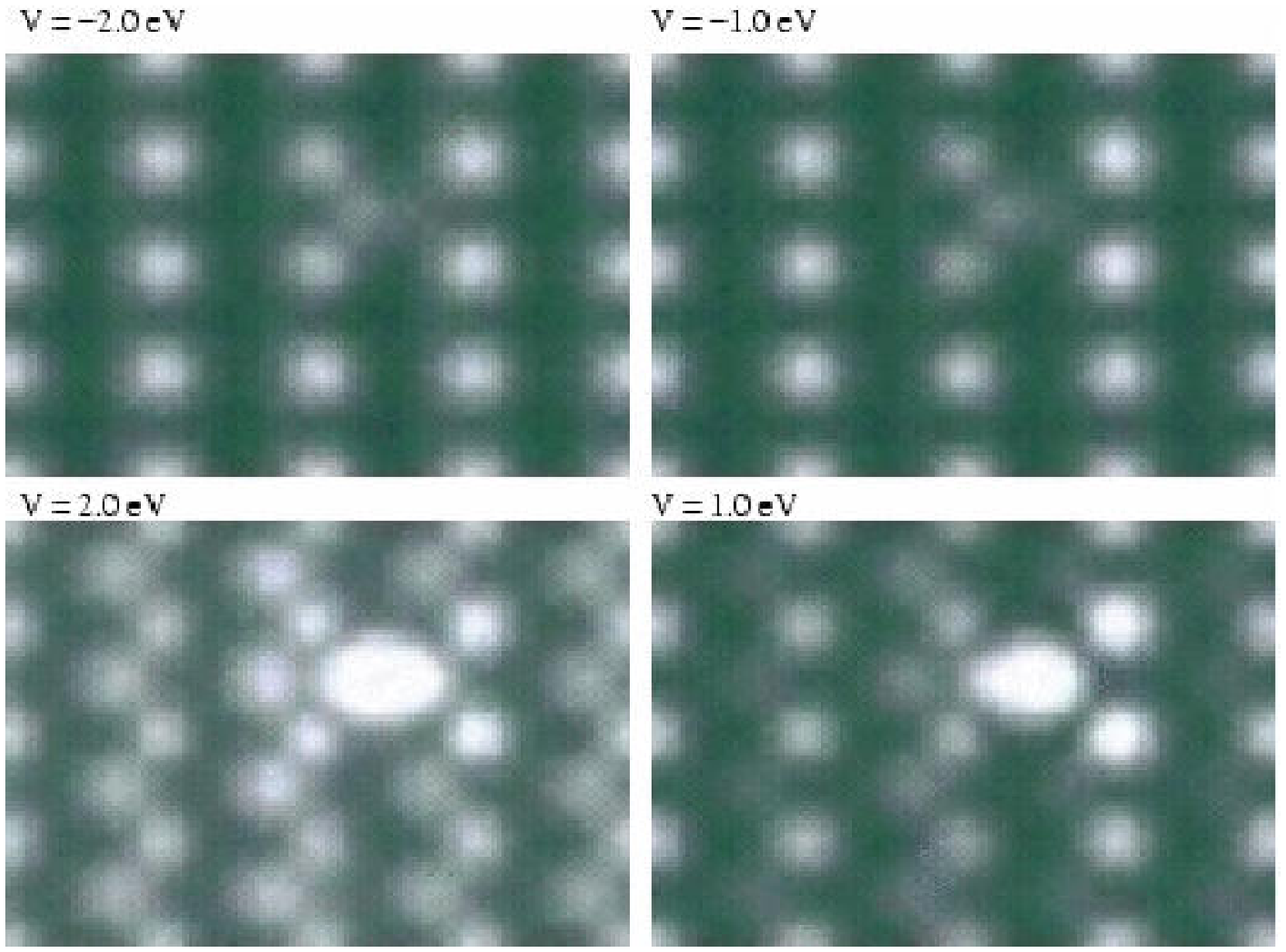}
\vspace{1cm}
\\Fig. 3
\end{figure}

\newpage
\clearpage



\begin{thebibliography}{99}

\bibitem{spintronic1} H. Ohno, Science 281, 281 (1998) 951.\vspace{1cm}
\bibitem{spintronic2} Y. Ohno, D.K. Young, B. Beschoten, F. Matsukura, H. Ohno, D.D.
Awschalom, Nature (London), 402 (1999) 790.\vspace{1cm}
\bibitem{spintronic3} H. Ohno, D. Chiba, F. Matsukura, T. Omiya, E. Abe, T.
Dietl, Y. Ohno, K. Ohtani, Nature (London), 408 (2000)
944.\vspace{1cm}
\bibitem{Spinbis} I. Malajovich, J.J. Berry, N. Samarth, D.D. Awschalom,
Nature (London), 411 (2001) 770.\vspace{1cm}
\bibitem{gamnas1} D. Chiba, K. Takamura, F. Matsukura, H. Ohno, Appl. Phys.
Lett., 82 (2003) 3020.\vspace{1cm}
\bibitem{PRLMN} K.W. Edmonds, P. Boguslawski,
K.Y. Wang, R.P. Campion, S.N. Novikov, N.R.S. Farley, B.L.
Gallagher, C.T. Foxon, M. Sawicki, T. Dietl, M.B. Nardelli, J.
Bernholc, Phys. Rev. Lett., 92 (2004) 37201.\vspace{1cm}

\bibitem{Modesti} S. Modesti and D. Furlanetto, private communication.\vspace{1cm}


\bibitem{PZ} D.M. Ceperly, B.J. Adler, Phys. Rev. Lett., 45
(1980) 566. \vspace{1cm}
\bibitem{PZ1} J. Perdew, A. Zunger, Phys. Rev. B, 23 (1981) 5048.\vspace{1cm}
\bibitem{pwscf} S. Baroni, A. Dal Corso, S. De
Gironcoli, P. Giannozzi, C. Cavazzoni,
http://www.pwscf.org.\vspace{1cm}
\bibitem{Vander} D.H. Vanderbilt, Phys. Rev. B, 41 (1990) 7892.\vspace{1cm}
\bibitem{forces} For the optimation of atomic positions we require
 Hellmann-Feynman forces smaller then 0.02 eV\AA$^{-1}$.\vspace{1cm}

\bibitem{Ters1} J. Tersoff, D. Hamann, Phys. Rev. Lett.,
50 (1983) 1998.\vspace{1cm}
\bibitem{Ters2} J. Tersoff, D. Hamann, Phys. Rev. B, 31
(1985) 805.\vspace{1cm}
\bibitem{STMGAAS} R.M. Feenstra, J.A. Stroscio, J. Tersoff, A.P. Fein,
 Phys. Rev. Lett.,
58 (1987) 1192.\vspace{1cm}
\bibitem{Duke} C.B. Duke, J. Vac. Sci. Technol., 1 (1983) 732.\vspace{1cm}
\bibitem{bulkint} J.Ma\v{s}ek, J.Kudrnovsk\'{y}, F.M\'{a}ca,
Phys. Rev. B, 67 (2003) 153203.\vspace{1cm}
\bibitem{condmat} J.X. Cao, X.G. Gong, R.Q. Wu, cond-mat/0503520.\vspace{1cm}


\end{thebibliography}
\end{document}